\documentclass[12pt,preprint]{aastex}

\shorttitle{The Cas A Progenitor}
\shortauthors{Young et al.}

\newcommand{\sol}{$M_\odot$\,}
\def \nuc#1#2{\relax\ifmmode{}^{#1}{\protect\text{#2}}\else${}^{#1}$#2\fi}

\begin{document}

\title{Constraints on the Progenitor of Cassiopeia A}

\author{Patrick A. Young\altaffilmark{1,2}, Chris L. Fryer\altaffilmark{1,3}, Aimee Hungerford\altaffilmark{1}, David Arnett\altaffilmark{2}, Gabriel Rockefeller\altaffilmark{1,3}, F. X. Timmes\altaffilmark{1}, Benedict Voit\altaffilmark{1,4}, Casey Meakin\altaffilmark{2}, and Kristoffer A. Eriksen\altaffilmark{1,2}}
\altaffiltext{1}{Theoretical Astrophysics, Los Alamos National Laboratories, Los Alamos, NM 87545}
\altaffiltext{2}{Steward Observatory, University of Arizona, 
 Tucson AZ 85721}
\altaffiltext{3}{Physics Dept., University of Arizona, 
 Tucson AZ 85721}
\altaffiltext{4}{Department of Computer Science, The University of Texas at Dallas, Richardson TX 75080}
\email{payoung@lanl.gov, fryer@lanl.gov, aimee@lanl.gov, darnett@as.arizona.edu, gaber@lanl.gov, fxt@lanl.gov, cmeakin@as.arizona.edu, keriksen@as.arizona.edu}

\begin{abstract}

We compare a suite of 3D explosion calculations and stellar models
incorporating advanced physics with observational constraints on the
progenitor of Cassiopeia A. We consider binary and single stars from
16 to 40 \sol with a range of explosion energies and geometries. The
parameter space allowed by observations of nitrogen rich high velocity
ejecta, ejecta mass, compact remnant mass, and $^{44}$Ti and $^{56}$Ni
abundances individually and as an ensemble is considered. A progenitor
of 15-25 \sol which loses its hydrogen envelope to a binary
interaction and undergoes an energetic explosion can match all the
observational constraints.

\end{abstract}


\section{INTRODUCTION}

Cassiopeia A is perhaps the best studied young Galactic supernova Remnant
(SNR). It is nearby (3.4 kpc) \citep{reed95} and young ($\sim 325$ yr)
\citep{thor01}. The wealth of data from ground-based observations in
the optical, IR, and radio and from space in the optical, x-ray, and
$\gamma$-ray allow us to study its morphology and composition in great
detail, and even observe its secular evolution. This remnant offers a
unique opportunity to place constraints on both the explosion and the
nature of the progenitor star. In this paper we will concentrate upon
the latter.

The identity of Cas A's progenitor has been the subject of tremendous
debate. Suggestions have ranged from a 16 \sol single star
\citep{che03} to various binary scenarios, to a Wolf-Rayet remnant of
a very massive ($\leq$ 60 \sol) precursor \citep{fb91}. We must
constrain the initial conditions far better if we are to use the large
amount of information obtained from the remnant to understand the
supernova explosion mechanism. It has also been suggested that Cas A
is related to Oxygen-rich SNRs in the Magellanic Clouds or to Type Ib
supernovae \citep[i.e.]{bl00}. Again we must be certain of
the progenitor's identity before accepting that Cas A is
representative of a large class of supernovae.

Most estimates for a progenitor mass are derived from one or few lines
of evidence. The lowest mass estimate of 16 \sol by \citet{che03} is
arrived at through self-similarity solutions for the explosion,
constrained by the positions of the forward and reverse shock and an
assumption of the structure of the circumstellar medium. The highest
masses estimates \citep[i.e.]{fb91} rest upon the lack of hydrogen in
the ejecta and the Wolf-Rayet-like properties of the pre-SN mass loss
deduced from the quasi-stationary flocculi. Though the highest mass
estimates in the literature have dwindled towards the low end of the
WR progenitor mass range (25-30 \sol), there is a lingering conception
of Cas A as a ``very massive star'' in the community. Other attempts have
been made to estimate a mass from nucleosynthetic products in the
remnant. These estimates vary wildly due to the enormous variation in
theoretical estimates of yields.

Most of these estimates contradict other lines of evidence than those
upon which they are based. For example, \citet{wil02} find that the
ejecta abundance ratios match the yield from a star that is 12 \sol at
the time of the explosion. From spectral line fits to the same data,
\citet{wil03} estimate a total ejecta mass of 2.2 \sol. If, as seems
likely, the compact remnant is a neutron star, the star must have been
$\sim$ 4 \sol at the time of explosion. Willingale et al. take care
not to make two contradictory claims about the progenitor, but the
case serves to illustrate the traps which may snare the unwary.

We attempt a different approach. We examine each of the major
observational constraints in turn, and compare them with progenitor
models. We then describe the parameter space allowed for the progenitor
if each of the constraints is accepted. Each constraint allows a wide
range of possibilities, but when the constraints are taken together,
almost all of those possibilities are eliminated. Section 2 describes
the progenitor models and explosion calculations. Section 3 evaluates
the major observational constraints and their potential level of
uncertainty. Section 4 evaluates our individual progenitor plus
explosion calculations in terms of the constraints. Section 5
describes the kinds of progenitors allowed by adopting the constraints
and examines the implications for our understanding of Cas A and
supernovae in general.

\section{PROGENITOR AND EXPLOSION CALCULATIONS}

\subsection{Progenitors}

We have produced four progenitor models as initial conditions for 3D
explosion calculations: two single stars, 40\sol and 23\sol, and two 
stars where the hydrogen envelope was removed during a red giant phase 
to mimic a common envelop evolution, 23\sol and 16\sol.  Although 
this sampling is sparse, it illustrates the effect of the observational 
constraints on the models and allow us to home in on the actual 
progenitor of Cas A.

All four models were produced with the TYCHO stellar evolution code
\citep{ya05}. In the absence of information on the composition of Cas
A's progenitor, we use the \citet{gs98} solar abundances to ease later
comparison with earlier calculations. The models are non-rotating and
include hydrodynamic mixing processes \citep{ya05,ymaf05}. Rotation is
a smaller effect than the hydrodynamic mixing in terms of core sizes,
and therefore masses of material at various stages of nuclear
processing, at core collapse. The magnitude of the effect of rotation
becomes comparable to that of wave driven mixing only for massive
stars rotating near breakup. The possibility remains that there is a
strong interaction between rotation and internal waves, but the
theoretical framework to evaluate this does not yet exist. In terms of
internal structure, we consider rotation to be a perturbation on the
existing hydrodynamic mixing smaller than the other uncertainties in
the study. The main area in which rotation may have an important
effect on this study is in enhancing the mass loss. Since mass loss in
uncertain to begin with, we discuss below the possible impact of
increased mass loss on our results. Angular momentum is also important
to some supernova mechanisms, such as the jet driven model, but we do
not use such a mechanism here. Explosion asymmetries which may be
induced by rotation are examined in a parametrized way. Mass loss uses
the prescriptions of \citet{kud89} for OB mass loss, \citet{bl95} for
red giant/supergiant mass loss, and \citet{ln03} for WR phases.

These progenitor models are an improvement over earlier studies, but
much work remains to be done in stellar physics. While our description
of the hydrodynamic mixing is unlikely to change qualitatively, it
will change in detail as simulations improve our understanding. A good
understanding of angular momentum transfer in stellar interiors is
lacking. Our picture of mass loss also has shortcomings, especially
for cool stars. The effects of waves on the URCA process and neutrino
cooling are not included, nor have wave-driven non-radial
perturbations been imposed on the initial conditions for the
explosion.

The 40 \sol star develops an instability during the late main sequence
which should lead to a Luminous Blue Variable (LBV) phase, so we
remove 1 \sol from the envelope as a conservative estimate for the
mass lost in eruptions. This model develops a core of He plus triple
$\alpha$ products with a maximum extent of 19 \sol and a very thin ($<
0.1$ \sol envelope of CNO burning products. Since N is rapidly
destroyed at triple $\alpha$ temperatures, the N rich material
disappears quickly as the thin CNO layer is removed by mass loss. The
core is subsequently eroded by mass loss as a WC and then WO star. The
final mass at collapse is 7.8 \sol.

The single 23 \sol star evolves normally as a red supergiant. The
final mass is 14.4 \sol, with a 5 \sol H-rich envelope. The main
uncertainty in this progenitor is the extent of red supergiant mass
loss. Mass loss predictions for cool stars are based on empirical
relations, not physical theories. Our rates are similar to those
observed, but a factor of two change could be admissible. This star
has a core of material with \={A} $\geq$ 16 of $\sim$6.5 \sol, with an
additional 3 \sol of material partially processed by triple $\alpha$
burning. The internal structure differs substantially from the 23
\sol progenitor with a ``binary event''.

The 23 \sol star with artificial binary evolution has its hydrogen
envelope removed when the star is at the base of the first ascent red
giant branch. At this point the radius exceeds 200 $R_{\odot}$, which
is larger than the separation of many massive star binaries.  The
convective envelope has not reached the hydrogen burning shell, so the
He core size and abundance profile of CNO products within the core are
not modified. We do not attempt to change the structure of the star
aside from removing the envelope.  We can place some weak constraints
on the nature of the companion assuming that the reason we do not see
it is that it merged with the primary star during this
common envelope phase.  A companion of $>$0.9 \sol could remove the
entire hydrogen envelope assuming standard common envelope evolution
and parameters \citep{FBB98}. At a separation of 200 $R_{\odot}$ a
companion $\gtrsim$ 2.6 \sol will not merge. The maximum mass for a
merger increases with decreasing separation. 

If the companion is lower mass and still in hydrogen burning, the
density contrast with the primary's core should be sufficiently large
that the mixing of material between stars will be limited to
Kelvin-Helmholtz instabilities at the interface with the accreted
material if the companion merges. If we also assume the entropy
barrier of the H burning shell limits the extent of any mixing process
driven by a non-merger, our results will change little. We assume by
necessity that the amount of mass accreted from the merger of a
companion is small compared to the mass of the primary. Mass loss
removes any residual H, and the star evolves as a WNL and WN until
core collapse. The final mass is 6.4 \sol, with a nitrogen-rich He
envelope.

The 16 \sol binary reaches a final mass of 5 \sol. At collapse there
is a residual envelope of  $\sim$ 0.01 \sol with 30\% hydrogen. A
further 0.5 \sol of material is completely H depleted but N
enriched. A significant fraction of this material does not have CNO
equilibrium abundances because of mixing of C/O rich material out from
the He shell convective zone.

\begin{figure}
\figurenum{1}
\plotone{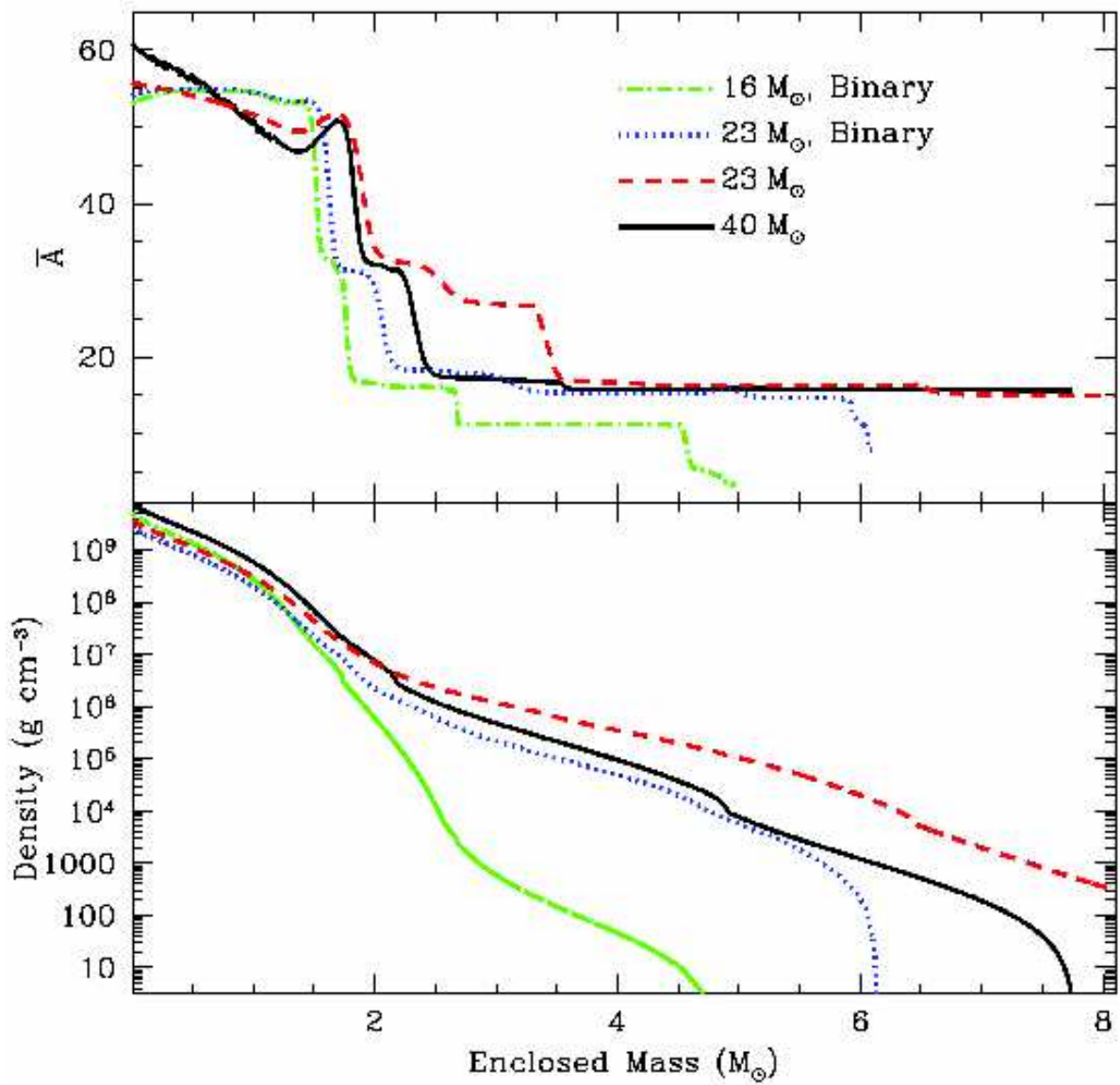}
\caption{Mean atomic weight \={A} (top) and density (bottom) vs. mass
coordinate for the four progenitor models. The extreme mass loss of
the 40 \sol WR and the binaries result in lower densities in the
oxygen dominated part of the core compared to the 23 \sol RSG. In the
Fe-peak dominated core the densities are higher because $Y_e$ is
lower.\label{fig1} }
\end{figure}

Figure~\ref{fig1} shows mean atomic weight \={A} (top) and density
(bottom) vs. mass coordinate for the four progenitor models. The
extreme mass loss of the 40 \sol WR and the binaries results in lower
densities in the oxygen dominated part of the core, compared to the 23
\sol RSG. In the Fe-peak dominated core, the models with higher
densities have lower $Y_e$. Most of the support in this region is
supplied by electron pressure, so with fewer electrons per nucleon,
the density must be higher to provide pressure support. Higher
densities out to large mass coordinate in the 23 \sol RSG will
significantly delay the explosion, and will result in a weaker
explosion for a given neutrino luminosity.
\clearpage
\clearpage
\subsection{Explosion Calculations}

We model the explosion of these stars using a two-step process: the
collapse and launch of the explosion modeled in 1-dimension and the
propagation of the shock through the star modeled in 3-dimensions.
The 1-dimensional code uses a coupled set of equations of state to
model the wide range of densities in the collapse phase
\citep[see][for details]{her94,fryer99} including a 14-element nuclear
network \citep{bth89} to follow the energy generation. This network
terminates at $^{56}$Ni and cannot follow neutron excess. A 512
element network was used in post-processing to derive more accurate
yields.  This code uses a flux-limited diffusion transport scheme for
3 species of neutrinos ($\nu_e,\bar{\nu}_e, {\rm \, and \,} \nu_x=\mu
{\rm \, and \,} \tau$ neutrinos) with a ``trapping radius'' defined as
the position where optical depth out of the star for a given neutrino
species is less than a given value.  For the simulations presented
here, this value is set to an optical depth of 0.1.  We artificially
raise the neutrino flux at this trapping radius to produce explosions
of varying energies.

After the launch of the shock, we cut out the neutron star (replacing
it with a hard reflective boundary and a gravitational potential
defined by its baryonic mass).  We continue to follow the explosion in
1-dimension until nuclear burning is finished (roughly 10-100\,s after
the launch of the explosion).  At this point, the output from the
1-dimensional simulation is mapped into our 3D explosion code: SNSPH
\citep{FRW05}.  We follow the explosion with this code to study the
mixing produced as the shock moves through the star.  The amount of
mixing both determines the final yield of the explosion and the
spatial position of the elements relative to each other.  It is this
mixing that (hopefully) causes the inversion of some of the iron and
silicon ejecta seen in the X-ray images \citep{Hwa04}.

As the focus of this paper is on the progenitor, we defer discussion
of the actual spatial distribution and detailed nucleosynthesis of the
ejecta to a later paper.  Instead, we concentrate on the remnant and ejecta
masses from our explosion calculations and bulk yields of $^{44}$Ti
and $^{56}$Ni from minimal post-processing of the explosive
nucleosynthesis.  We have run 13 explosion calculations, using our 4
progenitors with different explosion energies and asymmetries.  The
asymmetric explosion is a sinusoidally varying jet explosion with a
factor of 2 increase in the velocity along one axis \citep{HFW04}. The
explosion energy is defined as the kinetic energy of the ejecta at the
end of the simulation.  Table \ref{tab:exp-sims} shows the results
from this suite of simulations.  There are a few important trends in
these results.

First, note that the single-star progenitors (the 40 \sol star and
the single 23 \sol star) require much higher explosion energies to
avoid the formation of a black hole.  The compact cores of these
single stars are more difficult to explode.  It is likely that,
if they explode, they explode later with weaker explosions
\citep{Fryer99}.  Such weak explosions will always lead to a lot of
material falling back onto the compact remnant, forming a black hole.  
To avoid this fate, we must drive considerable energy into the 
gain region of the star.  The explosion then occurs roughly at 
the same time as the lower mass progenitors, but because the 
explosion must overcome a higher ram pressure, it is much stronger.
If we could show that the compact remnant from Cas A were a
neutron star, then this would constrain both the progenitor and its
explosion energy.  

The more energetic explosions have stronger shocks which extend the
explosion burning further out in the star.  This increases the yield
of many of the explosive burning products such as $^{56}$Ni and
$^{44}$Ti.  Upper limits on the $^{56}$Ni and a measurement of 
the $^{44}$Ti will also constrain our explosion energy and progenitor 
(\S~\ref{sec:cons-Ti44}).

Asymmetric explosions allow some wiggle room in these yield and
compact remnant constraints.  An asymmetric explosion of the same
energy as a symmetric explosion will eject more heavy elements, but
produce a larger compact remnant.  Where the explosion is weaker than
the symmetric explosion, the material is more likely to fall back and
increase the mass of the compact remnant.  Where the explosion is
stronger, there is increased burning and mixing.  Table
\ref{tab:exp-sims} shows the increased yield just from the increased 
mixing.  Asymmetries can dramatically change the heavy 
element yield.

\section{OBSERVATIONAL CONSTRAINTS ON PROGENITOR MODELS}

There are a variety of constraints on the possible progenitor arising
from observations of Cas A itself. Some of these are
considerably more stringent than others. In this section we examine
the nature and strength of the major constraints.  We then apply 
these constraints on our specific simulations in \S \ref{sec:obsvsim}, 
and then to general stars in \S \ref{sec:obsvstar}.

\subsection{Nitrogen Knots}

There are some four dozen observed high velocity nitrogen-rich knots in
Cas A \citep{fesen01}. As the supernova shock moves through the star,
the nitrogen layer is accelerated.  Because nitrogen is an
efficient coolant, the nitrogen ejecta quickly cools and forms knots.
These knots are observed to have high velocities of $\sim 9000 {\rm \ km
\ s^{-1}}$.  To achieve such high velocities, the nitrogen must have
been near the surface of the star (where the shock velocity is high)
when the star exploded.  

Furthermore, the material is nitrogen rich but hydrogen poor. The N to
H$\alpha$ flux ratios are tens of times solar with just three
exceptions. Most of the knots only have upper limits for H$\alpha$
flux.  In order to have N/H $\sim 30 \times$ solar, a typical limit
for nitrogen-rich knots, the material must have undergone CNO
processing until $>$90\% of the hydrogen was depleted, but not have
reached triple $\alpha$ burning temperatures \citep[and references
therein]{arnett96}. When the star exploded, the He core must have been
exposed, but not sufficiently eroded by mass loss to reach the C/O
rich region.

Thus any progenitor must fit into a very narrow end-state, where most 
of the hydrogen envelope has been removed, but the C/O-rich core can not  
be exposed; the nitrogen is still in the star.  The constraint 
on the progenitor is that it must have ended its life as a WN or 
WNL star. There is no other obvious way to produce these nitrogen 
knots, so we will take this constraint as solid.

\subsection{Ejecta Mass}
\label{sec:cons-mass}

It would be useful to have a direct measurement of the mass of the star
when it exploded. Though this is not possible, we can attempt an
inventory of the material associated with Cas A today.

We begin with the mass of the ejecta expelled in the
explosion. Estimating the ejecta mass is not straightforward. Previous
research has used two principal methods. The first combines optical
measurements of the positions of the forward and reverse shocks in the
remnant with kinematic arguments to determine a mass and explosion
energy \citep{che03,lh04}. The one dimensional nature of these
arguments in their present form and assumptions about the nature of
the circumstellar medium and distribution of explosion energy
introduce uncertainties into such estimates.  Such calculations
predict small ejecta masses in the 2-4 \sol range.

The second estimate uses observations of the x-ray emitting gas
\citep{wil02}. X-ray spectral line fitting combined with emission
models provides estimates of electron density and temperature, ion
temperature, composition, and emissivity, which can be used to
estimate the total amount of emitting material. The filling factor of
the material, the ratio of ion to electron temperature
$\frac{T_{ion}}{T_e}$, and the presence of a reservoir of material
which is at a temperature where emission is inefficient can all change
the mass estimate.  

Both methods of determining the ejecta mass are very model dependent,
but rely upon different sets of assumptions. The two methods both
arrive at similar ejecta masses (2-4 \sol).  Assuming that one method
did not bias the other, and the errors in the two techniques have not
led to the same incorrect answer, we can take this ejecta mass 
result as reasonably strong.

\subsection{The Compact Remnant}
\label{sec:cons-remn}

To truly estimate the total mass of the star prior to collapse, we
must also have an estimate of the compact remnant mass.  If the
compact remnant is a neutron star, we can set an upper limit to the
remnant mass equal to the maximum neutron star mass.  For most
equations of state, this maximum neutron star mass for a slowly
spinning neutron star is roughly 2.2\,M$_\odot$ in baryons
\citep[i.e.][]{mbs04}.  If the compact remnant is a black hole, its mass
is still weakly constrained by the nucleosynthetic yields (see \S
\ref{sec:cons-Ti44}).  Depending upon the level of asymmetry in the
explosion, the limit for a black hole could be as high as
5\,M$_\odot$, opening up a much wider range in Cas A progenitors.
Here we discuss the current evidence for the nature of the compact
remnant.

The compact remnant of Cas A was discovered in the first-light image
from the {\it Chandra} X-ray satellite \citep{tan99}.  Since this
time, observations from both the {\it Chandra} and XMM satellites have
sought to pin down the nature of this remnant \citep{chak01, mur02,
mti02}.  These observations, plus upper limits placed by observations
in other wavelengths, have begun to constrain the nature of this
compact remnant.

\citet{chak01} outlined the possible characteristics for the Cas A
compact remnant: classical pulsar, accreting neutron star, accreting
black hole, cooling neutron star, or anomalous X-ray pulsar/soft
gamma-ray repeater (AXP/SGR).  Due to the lack of a convincing
detection of pulsations \citep{mur02} and of any plerion, they
conclude that it is unlikely the remnant is a classical,
rapidly-spinning pulsar.  This justifies our use of a maximum neutron
star mass based on a slowly spinning neutron star.  The optical to
X-ray flux ratio strongly constrains accretion models. It is
inconsistent with disk accretion in a normal low mass x-ray binary
(LMXB), though accretion of SNR material may produce emission little
like normal binary disk accretion. Models of accretion onto a weakly
magnetized (B $< 10^9$\ G) neutron star can produce the correct x-ray
flux from the accretion boundary layer, but higher magnetic fields
either prevent boundary layer formation or result in strong pulsations
unless the NS is slowly rotating. The inner edge of a black hole
accretion disk has too large an emitting area for x-rays, and
advection dominated accretion flow (ADAF) and coronal models require
fine-tuning \citep{chak01}. It should be noted that the correct value
of the x-ray luminosity is dependent upon the interstellar
absorption. At present the amount of absorption is model dependent
\citep[see Table 2 of]{chak01}. \citet{pav00} were forced to invoke
extreme conditions to make a simple cooling model viable.  Despite Cas
A's X-ray luminosity being 3-10 times dimmer than the dimmest existing
AXP, the AXP/SGR scenario fits the data reasonably well and it has
become the favored remnant scenario for Cas A. This case has recently
been strengthened by optical and near IR limits from
\citet{fes05}. Observations with NICMOS on HST detect no R, J, or H
band counterpart to the compact object to magnitude limits of 28,
26.2, and 24.6, respectively.

Recently, \citet{kra05} have detected what they infer to be
infrared echos arising from a $2\times10^{46}$\,erg outburst from the
compact remnant.  Assuming this burst of non-periodic emission arises
from an SGR, this supports the idea that the remnant is an AXP/SGR.
It would be difficult to explain such an outburst from an accretion
scenario.

\begin{figure}
\figurenum{2}
\plotone{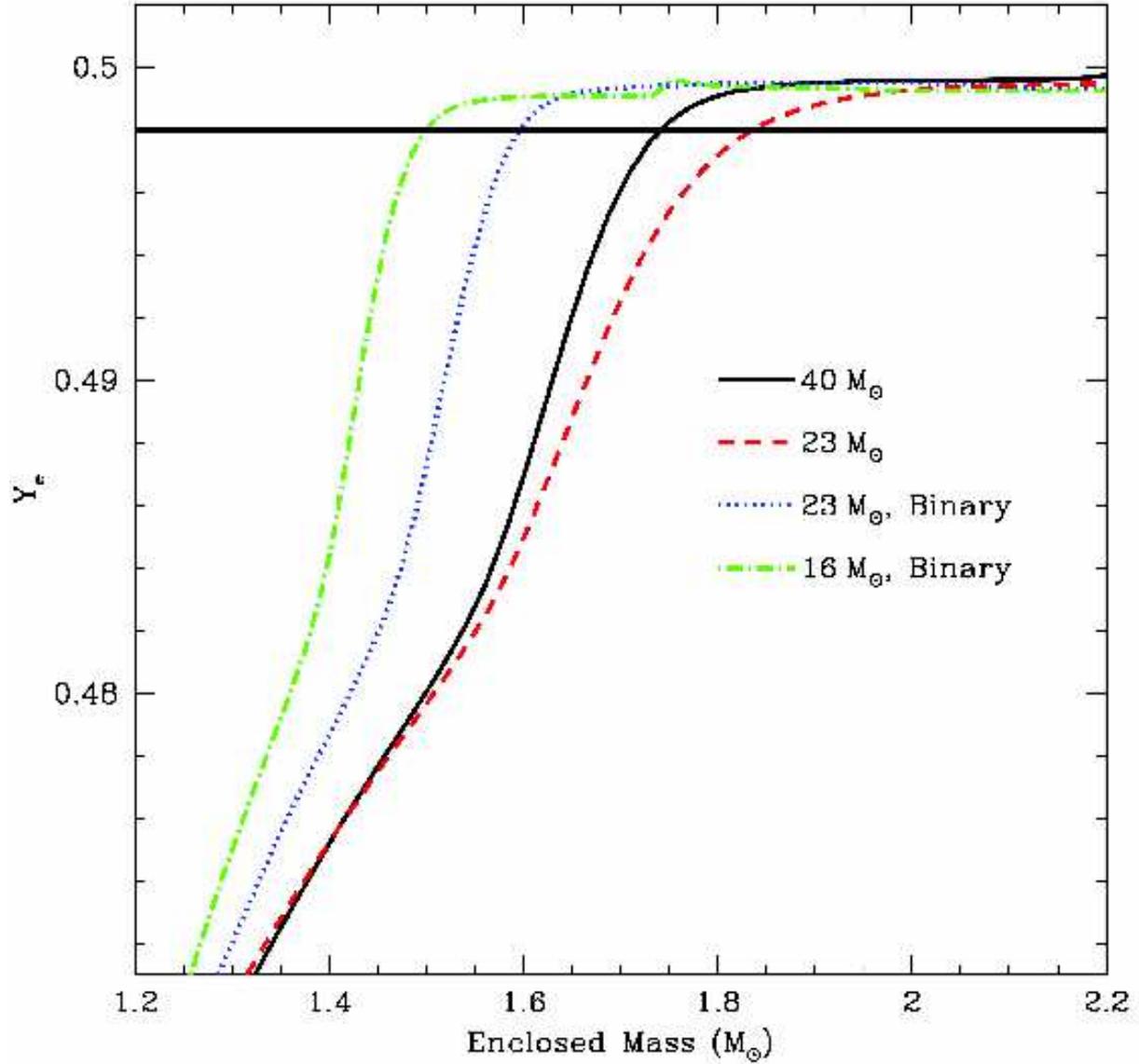}
\caption{$Y_e$ vs. enclosed mass for the four progenitor models. The
lower the mass during Si burning, the smaller the neutron enriched
core. The horizontal line indicates $Y_e = 0.498$, the approximate
electron fraction for the solar iron peak. Very little material more
neutron rich than this can be allowed to escape. This marks the
minimum mass of the compact remnant compatible with galactic chemical
evolution constraints. \label{fig-ye} }
\end{figure}

We can also place a lower limit on the neutron star mass. This limit
is stronger to the extent that Cas A is a typical supernova rather
than an odd event. It has long been known \citep[i.e.][]{hcaw74} that
the dominant nuclei in the e-process are a sensitive function of the
neutron excess $\eta$ (or equivalently the electron fraction $Y_e=
0.5(1-\eta)$ ). In order to reproduce the solar system isotopic
abundances of Fe and Ni, a $Y_e$ in the range of $1.5\times10^{-3}$ to
$3.0\times10^{-3}$ is required \citep[p270-272,321]{arnett96}. This
corresponds to $Y_e$ = 0.4992 to 0.4982. $^{56}$Fe is produced as
$^{56}$Ni $(Z=N=28)$ which decays to $^{56}$Fe through $^{56}$Co. The
most significant neutron-rich isotopes in the iron peak are $^{54}$Fe
$(Y_e = 0.485)$ and $^{58}$Ni $(Y_e = 0.483 )$.  Requiring that they
are produced in (at most) their solar system value gives $Y_e \ge
0.4982$. More neutron-rich material must not have been ejected in
significant amounts. Figure~\ref{fig-ye} shows the electron fraction
$Y_e$ of the inner cores of our four progenitors. The matter interior
to 1.75 \sol\footnote[1]{These are baryonic masses. Gravitational masses
(measured in binary systems) are typically $\sim$10\% less massive
than baryonic masses. This mass deficit represent the gravitational
binding energy of the object} is too neutron-rich for the most massive
progenitors, but this limit is relaxed to 1.5 \sol or less for the lower mass
progenitors.  Although one could argue that the neutrinos streaming
out of the proto-neutron star can reset $Y_e$, this is only true for
matter very near the proto-neutron star surface. If Cas A is a
peculiar object, such as the product of common envelope evolution and
not a typical supernova, the $Y_e$ argument is weakened to the extent
that such objects rarely contribute to the ensemble of nucleosynthesis
sources from which the solar system was derived. Very non-solar
abundances relative to Fe might then be allowed, and the argument for
the minimum mass for the neutron star remnant no longer holds.

\clearpage
\clearpage
Although we will consider a black hole remnant in this study, we note
that the evidence is gradually increasing in support of a highly-magnetized,
slowly-rotating, neutron star remnant for this object.  We thus believe that
the remnant mass is most likely below 2.2M$_\odot$.

\subsection{${\rm ^{44}Ti \, and \, ^{56}Ni}$}
\label{sec:cons-Ti44}
As a final constraint, we consider the nucleosynthetic yields of both
radioactive titanium ($^{44}$Ti $\rightarrow$ $^{44}$Sc $\rightarrow$
$^{44}$Ca) and radioactive nickel ($^{56}$Ni $\rightarrow$ $^{56}$Co
$\rightarrow$ $^{56}$Fe).  Nuclear decay lines from $^{44}$Ca (1157
keV) and $^{44}$Sc (67.9 and 78.4 keV) have been solidly detected
towards Cas A with the CGRO COMPTEL and BeppoSAX PDS instruments,
respectively \citep{iyu94,iyu97,vin01}.  These detections, when taken
together, yield a line flux of $(2.5 \pm 1.0) \times 10^{-5}$photons
cm$^{-2}$ s$^{-1}$, which implies an initially synthesized mass of
$^{44}$Ti in the range of $(0.8-2.5) \times 10^{-4}$
M$_\odot$. Preliminary observations of the 68 keV line by ISGRI on
INTEGRAL are consistent with the COMPTEL and BeppoSAX results
\citep{vink05}. This mass of ejected $^{44}$Ti is generally thought to
be abnormally large (or at best, marginally consistent) by comparison
with spherical explosion models \citep{tim96}.

To complicate the matter further, \citet{tim96} argued that spherical
explosion models with an ejected mass of $\sim 10^{-4}$ M$_\odot$ in
$^{44}$Ti would imply an ejected $^{56}$Ni mass of at least 0.05
M$_\odot$.  At the distance of Cas A, this should have resulted in a
very bright supernova, yet historical records of the time show no such
entry. \citet{nag98} point out that the $^{44}$Ti/$^{56}$Ni ratio can
vary substantially due to asymmetries in the explosion. We also find
that the ration varies with the degree of asymmetry, and also with
fallback and the initial composiution and thermodynamic trajectory of
the material involved in $\alpha$-rich freezeout. It has been
suggested that the supernova associated with Cas A SNR was observed
and recorded by J. Flamsteed in 1680 \citep{ash80} at a visual
magnitude of 6.  Unlike the $^{44}$Ti, we have no solid constraints
for the synthesized $^{56}$Ni mass and must construct upper and lower
limits from indirect evidence.  In the remainder of this section, we
consider constraints on the $^{56}$Ni mass using two independent
techniques.  The first is based on an inventory of $^{56}$Fe (the
stable daughter product of $^{56}$Ni) in the present day remnant,
while the second relies on the absence of widespread reportage of the
supernova event in astronomical records of the time.

A simple inventory of the X-ray emitting iron mass in Cas A has been 
presented by \citet{wil03}.  They report an iron mass of 
0.058~M$_\odot$ and argue convincingly that most of this iron originated
in the supernova ejecta.  With reasonable accuracy, we can assume this iron
is comprised primarily of the $^{56}$Fe isotope, which gives
a relatively solid lower limit to the mass of radioactive $^{56}$Ni that must
have been synthesized and ejected by the explosion.  One could, in principle,
attain an upper mass limit by assuming the mass interior to the reverse shock 
is made entirely of $^{56}$Fe and then add the X-ray visible component to 
this number.  However, as \citet{Hwa04} discuss, it is possible for there to 
be iron already hit by the reverse shock that we can no longer detect.
As such, a solid constraint for the upper mass limit of $^{56}$Ni poses a
larger challenge and we turn instead to a constraint based on analytical
lightcurve models.

Given the lack of widespread reportage of the event, we assume that the 
peak, visual magnitude was fainter than m$_v \sim$ 3 
\citep{das93}.  For the general case, this includes the effect of 
visual extinction present at the time of the explosion.  From this
we can write down a simple expression for the absolute magnitude ($M_{peak}$) 
of the supernova at peak luminosity:

\begin{equation}
M_{peak} = 3 - A_v - 5{\mathrm{log}}\left(\frac{d}{10}\right)
\end{equation}

\noindent where $A_v$ is the magnitudes of visual extinction and $d$ is the 
distance to Cas A (3.4~kpc).

Using the sun (1 L$_\odot=3.83\times10^{33}$erg s$^{-1}$ and absolute
magnitude of 4.83) as a standard, we can relate $M_{peak}$ to a peak
luminosity of the supernova in units of solar luminosity.

\begin{equation}
L_{peak} = 10^{(\frac{M_{peak} - 4.83}{-2.5})} L_\odot
\end{equation}

This leaves us with a maximum luminosity for the Cas A supernova in order to 
avoid widespread reportage.  Our goal now is to derive through reasonable 
assumptions, the corresponding maximum mass of $^{56}$Ni.  Given the constraint
on the ejecta mass, it is very likely that the 4$-$6 M$_\odot$ object that
went supernova was initially of very small radius (as opposed to the standard
red, supergiant progenitor.)  With the assumption of small initial radius, we 
can employ a simple analytic model relating the supernova luminosity to its
initial mass of $^{56}$Ni (as described in \citet{arn82} and \citet{pin00}.) In
essence, this model equates the peak luminosity with the instantaneous energy
deposition from decay processes.

\begin{equation}
L_{peak}= M_{Ni}\Theta(t_{peak})\Lambda(t_{peak})
\end{equation}

\noindent where $t_{peak}$ is the time at peak luminosity, $M_{Ni}$ is the 
mass in grams of $^{56}$Ni synthesized,
$\Lambda$ is essentially one minus the energy escape fraction of
$\gamma$-rays and $\Theta$ is the instantaneous
energy decay rate for the radioactive $^{56}$Ni and $^{56}$Co.
$\Theta$ has units of erg~s$^{-1}$~g$^{-1}$ and is given by

\begin{equation}
\Theta(t) = \frac{N_{av}}{56}\left(\frac{E_{Ni}}{\tau_{Ni}}e^{-t/\tau_{Ni}} + 
\frac{E_{Co}}{\tau_{Co}-\tau_{Ni}}
\left(e^{-t/\tau_{Co}}-e^{-t/\tau_{Ni}}\right)\right) 
\end{equation}

\noindent where $N_{av}$ is Avagadro's number, $\tau_{Ni} = 7.6\times10^5$~s,
$\tau_{Co} = 9.6\times10^6$~s, $E_{Ni} = 1.73$~MeV and $E_{Co} = 3.69$~MeV 
are the mean decay lifetimes and energy per decay for $^{56}$Ni and $^{56}$Co 
respectively.  A quantitative value for this upper mass limit of $^{56}$Ni
requires that we specify $t_{peak}$ (the rise time to the peak of the light 
curve), $\Lambda(t_{peak})$ (the fraction of decay energy deposited at peak) 
and $A_v$ (the magnitudes of visual extinction toward Cas A.)

We choose $t_{peak}$ to be 18 days, which is a fairly typical rise
time for Type~Ib supernova lightcurves \citep{fil97}. We adopt a value
of 0.9 for $\Lambda(t_{peak})$ which is likely an underestimate.
Since our intent here is to derive an upper limit for $M_{Ni}$, erring
on the side of too little deposition is preferred.  In order to check
the validity of this assumption, we have run a $\gamma$-ray transport
simulation (with the Maverick code; details in \citet{HFW04}) using
our 16 and 40~M$_\odot$ progenitor models as input.  The simulation
results give a decay energy deposition of roughly 95\% for the 16\sol
and 100\% for the 40\sol at $t$ = 18.1 days. In order to reduce the
deposition, the material must be more optically thin, which requires
either a lower ejecta mass or very high expansion velocities. As such,
the 16\sol model is our best case for the escapse of $\gamma$-rays,
which supports our claim that 90\% deposition is a cautious lower
limit.

Current estimates for visual extinction values toward Cas A range from
4-8 magnitudes (\citep{pei71,tro85,pre95} and reviews by \citet{har97}
and \citet{die98}). There has been discussion in the literature
regarding the presence of a dust cloud at the time of the explosion,
which has since been evaporated by passage of the supernova shock.
Given the estimates for ejecta mass and swept-up mass in
\citet{wil03}, any dust condensation which was present at the
explosion and blasted away before the present epoch, was likely formed
out of the wind material during the star's giant phase evolution.
Typical extinction measurements toward WN stars possess visual
extinctions which range all the way down to $A_v < 1$, suggesting that
most of the extinction arises from the Galactic ISM rather than an
intrinsic source \citep{vdh01}. Also, CO measurements show that Cas A
lies near two giant molecular clouds (GMCs) in the Perseus arm, and
that extinction in the region will be large compared to a stellar dust
contribution \citep{uut00}.  This suggests that the extinction we see
today is likely to be similar to that obscuring the explosion itself.
If we adopt the larger end of this range (taken with $\Lambda = 0.9$
and $t_{peak} = 18$~days), we get a maximum mass for $^{56}$Ni of
0.2~M$_\odot$.  

\begin{figure}
\figurenum{3}
\epsscale{.9}
\plotone{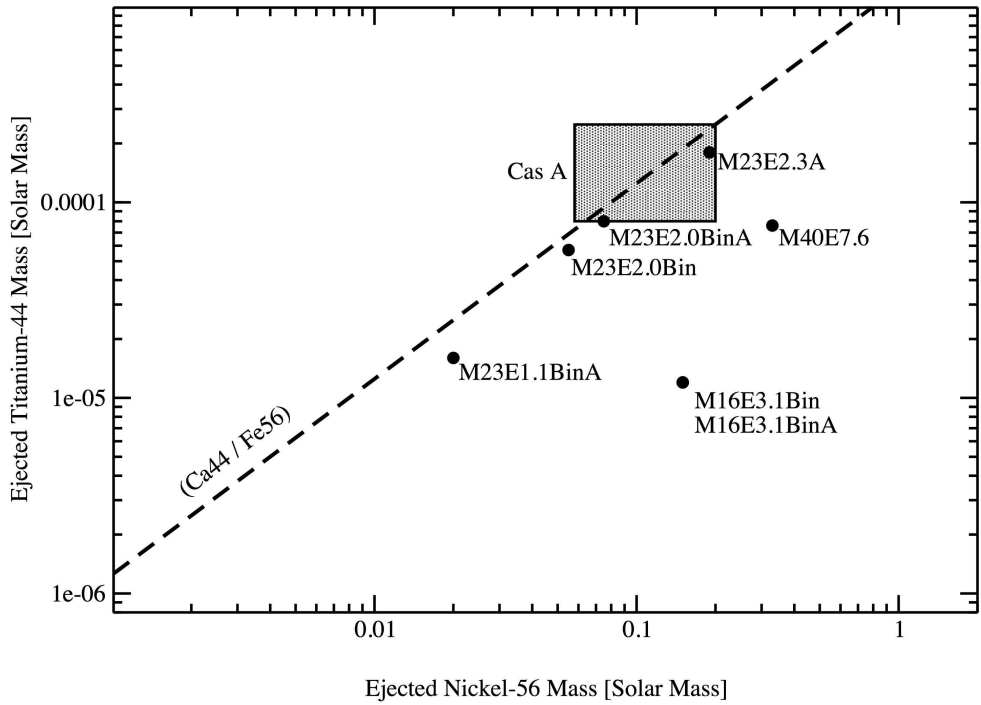}
\epsscale{1}
\caption{ejected mass of $^{44}$Ti against $^{56}$Ni. The range
allowed by the observations is shown by the stippled box. Dots are our
explosion calculations. Model M23E2.3A (2.3 foe asymmetric explosion
of 23 \sol non-binary) rests in the upper right of the
box. M23E2.0BinA (2.0 foe asymmetric explosion of 23 \sol binary)
falls in the lower left. The dotted line shows the solar ratio of the
principal decay products, $^{44}$Ca and $^{56}$Fe.\label{nucfig} }
\end{figure}

Figure~\ref{nucfig} plots ejected mass of $^{44}$Ti against
$^{56}$Ni. The dotted line shows the solar ratio of $^{44}$Ca to
$^{56}$Fe, the principal decay products. The range allowed by the
observations is shown by the stippled box. Our explosion calculations
are shown by the dots. Model M23E2.3A (2.3 foe asymmetric explosion of
23 \sol non-binary) rests in the upper right of the box. M23E2.0BinA
(2.0 foe asymmetric explosion of 23 \sol binary) falls in the lower
left. The asymmetric and symmetric explosions of the 2.0 f.o.e. 23\sol
binary and the 3.1 f.o.e. 16\sol binary have the same yields. This is
because there is little fallback. The differences in abundance ratios
in our other asymmetric models arise from differential fallback; some
material gets out that would not escape in a symmetric
explosion. Following the burning in 3D could change this result.

Clearly, the mass of $^{56}$Ni is subject to uncertainty, and is, by
far, the least stringent of our constraints. In addition, the
predicted mass of both $^{56}$Ni and $^{44}$Ti are both very sensitive
to the details of the explosion mechanism and
calculation. Figure~\ref{nucfig} shows that it is possible to produce
reasonable abundances from an otherwise promising progenitor, but
should not be taken to mean more than that. We do not believe these
abundances can be used to distinguish between progenitor models at
this stage.
\clearpage
\clearpage

\section{Observations versus Simulations}
\label{sec:obsvsim}

By reviewing the fate of our specific models, we can build up our
intuition on what is (and is not) important in producing a viable Cas A
progenitor.  Table \ref{tab:obsvsim} shows the status of our models
when compared against these observational constraints. We study 
each progenitor in turn.

Mass loss in the 40 \sol progenitor removes all of the nitrogen rich
material and a significant amount of C/O rich material. It cannot
explain the nitrogen-rich knots.  Also, because its final mass is over
6 \sol, it can not satisfy both the ejecta and remnant mass
constraints. Increased mass loss during the LBV or WR phase could
reduce the final mass to an acceptable value, but it would still fail
to explain the N-rich knots. We ran a single very energetic explosion
to fit the remnant mass. This explosion produced somewhat too much
$^{56}$Ni.  A weaker explosion could fix this yield, but such a
progenitor can never match the nitrogen-rich knots nor the total final
mass constraints.

The single 23 \sol progenitor suffers the same fate as the 40 \sol
star: it can not match either the nitrogen-rich knots or total final
mass constraints. In this case the star retains a large hydrogen
envelope. Increased mass loss which could strip the H envelope would
have to occur as a luminous red supergiant. At this point the
convective envelope has penetrated deeply into the star. The He
burning convective core has also had more time to grow. Only a small
($\sim 0.1$\sol) layer of pure CNO ashes remains between the
convectively homogenized H-rich envelope and the C/O rich He burning
shell. This layer has a mass coordinate of $>$8 \sol, depending on
details of the mass loss history. We conclude that additional mass
loss as an RSG would be unlikely to leave a He/N surface layer. We
also notice another effect: to produce a neutron star, we almost
invariably eject too much $^{56}$Ni.

The binary 23 \sol progenitor can marginally match all the data.  It
requires a strong enough explosion to make a neutron star, but if the
explosion is strong enough, it appears that both the $^{44}$Ti and
$^{56}$Ni yields both lie within our set of constraints.  Note that
although the $2 \times 10^{51}$\,erg explosion with a mild asymmetry
can explain everything, the compact remnant is on the massive end.  A
better fit would probably arise from a slightly more energetic
explosion. 

The binary 16 \sol progenitor can also marginally match all of the
data except for the $^{44}$Ti. It fits nicely within the total mass
constraints, but has difficulty reproducing the yields. A strong
explosion produces enough $^{56}$Ni but too little $^{44}$Ti, while a
weak one produces too little of both $^{44}$Ti and $^{56}$Ni. It would
appear that a binary model somewhere between 16 and 23 \sol with a
moderately energetic explosion could fit the constraints well. It is
worth noting, however, that there are significant uncertainties in the
predicted yields.

\section{Possible Progenitors}
\label{sec:obsvstar}

In this section we present a series of options for the progenitor's
identity as we descend the decision tree of constraints. The strongest
of the constraints is the presence of fast-moving Nitrogen-rich,
hydrogen-poor ejecta. At the time of the explosion, the star's surface
composition was mostly CNO processed ash. There does not appear to be
any way to avoid this conclusion, and all of our options will follow
from it. We do not separately consider the nucleosynthesis
constraints. They are weaker observationally than the other
constraints, and the nucleosynthesis is far too model dependent to be
a good discriminator between otherwise viable alternatives.

A high mass Wolf-Rayet progenitor (such as our 40 \sol progenitor)
will uncover its C/O core before core-collapse.  The observation of
nitrogen-rich knots excludes such a progenitor.  Unfortunately,
uncertainties in LBV and Wolf-Rayet mass-loss rates make it difficult
for us to place an upper limit on the progenitor mass due to this
constraint.  Our conservative estimate for this limit is 28-30 \sol.

For single stars, we can also place a lower limit on the progenitor
mass.  Low mass stars will not lose their hydrogen envelope by core
collapse. For example, by the time the 23 \sol single-star model
reaches the red giant branch it has lost less than 3 \sol, leaving
almost 10 \sol still in the hydrogen rich envelope. The star does lose
a further 6 solar masses, and it is conceivable that our mass loss
rates may be off by a factor of two for red supergiants. However, we
know that some stars in the 15 -25 \sol mass range retain a hydrogen
envelope from the example of SN87A. Even if the mass loss is sufficient
to strip the envelope, most of it must occur when the star is a
luminous red supergiant, meaning that the convective envelope has
penetrated deeply into the star, and the He burning convective core is
large. Only a small ($\sim 0.1$\sol) layer of pure CNO ashes remains
between the H-rich envelope and the C/O rich He burning shell. It
would require considerable fine tuning to time core collapse to
coincide with the short period during which such a star would have CNO
ash-rich surface layers. In addition, this layer has a mass coordinate
of $>$8 \sol for a 23 \sol progenitor, well outside the predicted size
of the Cas A progenitor.  For binary systems, this lower limit on the
mass is removed, since the H envelope can be removed early.

The first branching point for progenitor options rests upon the
estimates of total ejecta mass.  From Section~\ref{sec:cons-mass}
above, the ejecta mass has been estimated from similarity solutions
constrained by the positions of the optical forward and reverse shocks
and from x-ray spectral fits.  Both of these methods produce similar
values for the ejecta mass (2-4 \sol), but both are also model
dependent. As we show below, relaxing this constraint fundamentally
changes the parameter space available for a progenitor. Detailed
modeling and additional observations which would provide a more secure
and complete determination of the ejecta mass are desirable.

\subsection{High Ejecta Mass}

If we take the view that the mass determinations quoted above
represent a lower limit on some incompletely sampled mass of ejecta,
we are given more options for a progenitor. In this case it is
possible to admit a high mass star ($>$ 30 \sol)which has evolved into a
WN or WNL star through single star evolution. Such a star could
produce a neutron star remnant with a sufficiently energetic
explosion, but would produce at least six solar masses of O-rich
ejecta. Alternatively it could produce less ejecta and a black hole
remnant with a weaker explosion. We may think of this as a ``normal''
Type Ib SN.

\subsection{Low Ejecta Mass}

If we accept that the current inventories of ejecta mass are
substantially complete and accurate, we are left with two options for
a progenitor which depend upon the nature of the compact object. The
sum of mass in the ejecta and compact remnant give us the total mass
of the star at core collapse. We present two options, one for a
neutron star compact remnant and one for a black hole.

\subsubsection{Black Hole Remnant}

This scenario is the least constraining of our final two options. A
somewhat arbitrary amount of mass can be incorporated into a black
hole, so we have only weak constraints on the mass of the star at
collapse. Much as in the high ejecta mass model, we are allowed a more
massive progenitor which can produce a WN or WNL star from single star
evolution. In this case a weak explosion with a large amount of
fallback produces$\sim$ 2 \sol of ejecta and a $\lesssim$ 6 \sol black
hole, again allowing progenitors $<$ 30 \sol. Such weak explosions
would have difficulty producing enough $^{44}$Ti and $^{56}$Ni.

\subsubsection{Neutron Star Remnant}

If we accept all of our observational constraints to this point at
face value we arrive at this scenario. Current work on neutron star
equations of state place an upper limit on neutron star masses of
$\sim$ 2.2 \sol. The neutron star plus the observational estimates of
ejecta mass give us a total mass at core collapse of $\sim$ 4-6 \sol.

The very low mass at collapse, combined with the N-rich, H-poor
surface composition, imply a unique progenitor for Cas A. Massive
Wolf-Rayet progenitors have triple $\alpha$ processed cores which are
too large. Mass loss would uncover C/O rich material well before
reducing the core to 6 \sol. A lower mass star which retains a
hydrogen envelope until the RSG stage will mix the CNO processed
material with H-rich envelope material during dredge-up. Only a thin
layer of CNO ash at the hydrogen burning shell remains, leaving a
problem of fine-tuned timing if a star with such a small core can
remove its entire hydrogen envelope at all.

We require a scenario in which a relatively low mass star ($<$ 25 \sol),
which would not ordinarily produce a WN star, loses all or most of its
envelope early in its post main sequence evolution. Wolf Rayet mass
loss is then sufficient to reduce the star to its final small mass
while retaining a CNO processed outer layer. The immediately obvious
way to achieve this is to invoke common envelope evolution in a
binary. 

Our 23 \sol progenitor with ``binary envelope ejection'' can produce a
massive neutron star and 4 \sol of ejecta with a reasonable explosion
energy. The total core mass at explosion of 6.4 \sol puts it at the
massive end of the range allowed by the observational ejecta mass
determinations and theoretical neutron star equations of state. An
explosion energy above $\sim 2 \times 10^{51}$\,erg would be able to
explain all our constraints at the upper mass extreme of our
constraints.  A lower mass star ($\sim$ 16 \sol) which undergoes a
common envelope phase resides more comfortably with the parameter
space, with a final core mass of 5 \sol, though it has difficulty
producing the correct amount of $^{44}$Ti and $^{56}$Ni.  Details of
the envelope ejection and subsequent mass loss can shift these mass
estimates slightly.

We are left with a low mass (15$<M<$25 \sol) binary progenitor. Chance
would appear to have made our most easily studied young supernova
remnant a peculiar member of its class, rather than a truly
representative Rosetta stone. This very peculiarity however, probably
gives us much more insight into the progenitor and the nature of the
explosion.

\section{Implications}

The findings of this study have broader implications for both Cas A in
particular and supernovae in general. If we accept all of the
available observational constraints, the progenitor of Cas A turns out
to be atypical. A low mass star masquerading as a Wolf Rayet after
losing its envelope to a binary interaction is not an ideal model for
``normal'' supernovae, but may be more common than we think, given the
high binary frequency of massive stars. On the other hand, its oddness
may help to pin down the nature of the progenitor much better than
would otherwise be possible. This could remove some of the
uncertainties normally unavoidable in modeling the explosion, making
simulations a much more powerful tool for understanding the explosion
mechanism. We have laid out arguments for further refining the
constraints by both theoretical and observational means.

If the progenitor to Cas A was indeed a binary, then we are faced with
new issues. We would expect the binary interaction to give rise to
significant asymmetries in the circumstellar medium which should be
imprinted on the structure of the remnant. While Cas A shows marked
asymmetry in the northeast and southwest jet-like features, these
appear to be associated with the explosion and not the interaction
with the surrounding medium. The density of fast moving knots appears
to be lower in the north and south edges of the remnant
\citep{fesen01}, and there is some evidence of interaction with dense
material at the edge \citep[][i.e.]{fesen01,hines04}, but there is no
glaring evidence of bipolar structure. 

Asymmetries associated with binary interaction can show up in two
places. The first is the post-envelope ejection fast wind, which
is associated with the quasi-stationary flocculi (QSFs)
\citep{fb91}. There are three reasons why we might not see evidence
of asymmetry in this wind: 1) Cas A was not a binary at all; 2) an
accident of geometry placed our line of sight along an axis of
symmetry; 3) the companion star merged with the progenitor during the
common envelope phase. 

The second place where a binary would generate asymmetry is where the
expanding remnant runs into the ejected envelope. If the shock has
impacted this material we should see the emission, so reason two above
does not help us. Two simple scenarios are capable of addressing this
problem. Let us consider the mass swept up by the ISM. 
\begin{eqnarray}
\dot{M} \simeq 4\pi r^2v \rho_{ISM} \nonumber \\
\rho_{ISM} \simeq n_H/A \nonumber \\
\case{M}{M_{\odot}} \simeq \case{4\pi}{3}(vt)^2vt\case{n_H}{M_{\odot}A}  \nonumber \\
\case{M}{M_{\odot}} \simeq 13n_H(\case{v}{50{\rm km\ s^{-1}}})^3(\case{t}{10^5{\rm yr}})^3  
\end{eqnarray}
where $\case{M}{M_{\odot}}$ is the mass swept up, $n_H $ is the number
density of hydrogen, and $A$ is Avogadro's number.

Assuming an ejection velocity for the envelope of 50 ${\rm km\
s^{-1}}$ at $10^5$ yr before the explosion, the ejected material will
have reached 5 pc by the time the star exploded, if the surrounding
area already has $n_H \ll 1$, i.e. in a bubble swept out by an OB
association. The age and velocity both represent conservative
estimates for a star on the first ascent RGB, so it is quite possible
that the SNR simply has not yet encountered the envelope in this
scenario. At the opposite extreme, the envelope could have impacted a
very dense ISM, for example a molecular cloud, at an early time. The
kinetic energy of the envelope would be very small compared to the
inertia of the ISM, and would have imposed very little asymmetry. Any
forward shock would have dissipated quickly, and the reverse shock
would have long since propagated back to the origin, so we would see
no obvious evidence of that interaction at the present time. In this
scenario the SNR may already have propagated through the WR wind
bubble and be interacting with a combined envelope/ISM. Estimates of
the mass swept up by the remnant and indications of an interaction
with a molecular cloud tend to support the latter idea, but this is an
area which would benefit from additional observational scrutiny along
with multi-D simulation of the interaction between winds, SNRs, and
the ISM.

Detailed calculations of nucleosynthesis during the explosion will be
discussed in a subsequent paper, but this work already demonstrates
that estimates of yields are subject to severe uncertainties. We note
a few of these here as a prelude to the detailed calculations. Not
surprisingly, mixing of material during the explosion can drastically
change the yield of species produced deep in the star. A 1D explosion
with a mass cut cannot predict yields correctly. For example, a 3D
code will give larger Ti and Ni yields than 1D. This can mean the
difference between several hundredths of a solar mass of Ni and none
at all. 

Nuclear processing and adjustments to statistical equilibrium can
continue for tens of seconds. We find that several effects can result
in $1-2$ orders of magnitude changes in the abundance of species such
as $^{44}$Ti. For example, it is common to follow nuclear burning for
$\sim 5$s after the explosion. In these calculations the time to
completion of the freeze-out process was of order tens of seconds. Not
surpirsingly, the $\alpha$ network used in the explosion calculation
produces a much larger yield of $Z=N$ species (most notably $^{44}$Ti
and $^{56}$Ni) than the larger post-processing network. We also
discovered that the choice of time at which we cut out the neutron
star and replaced it with a hard inner boundary had an effect of
similar magnitude on the yields. The choice of explosion energy, which
is arbitrary so long as it doesn't contradict observational
determinations of the energetics, produces large changes, as does any
imposed asymmetry. We are forced to the conclusion that a much larger
and more thorough effort must be made to understand the theoretical
uncertainties in the yields before the observed yields can be used as
a meaningful constraint.

The primary location of Ni production is also debatable. It is common
in the observational literature to state that $^{56}$Ni is produced by
$\alpha-$rich freezeout of material from the progenitor's Si
shell. This is not at all clear. One zone nucleosynthesis models
starting with compositions taken from our progenitor models show a
wide variety of final compositions which depend sensitively upon both
the initial entropy and the thermodynamic trajectory of the
material. Often these compositions are not dominated by $^{56}$Ni; a
very small neutron excess is sufficient to poison the
freezeout. Because of the importance of even a small neutron excess,
substantial post processing is necessary. Results from an alpha
network cannot be taken at face value. It is common to ``reset'' $Y_e$
in the progenitor to remove the neutron excess. Neutrinos may be able
to reset $Y_e$, but much of the material in which neutrino deposition
could occur falls back onto the compact object in our simulations. It
may be that most of the Z=N material is produced by alpha rich
freezeout from the oxygen shell. The fate of the Si shell must be
calculated in detail for a given specific case. A great deal of
systematic theoretical work must be done before yields from an
explosion can be considered reliable.

The results of the explosions are very sensitive to the choice of
progenitor models. Mixing, mass loss, neutrino processes,
non-spherical displacements and perturbations, and angular momentum
can all influence the explosion, and all are imperfectly to poorly
understood. Aspects of the progenitor evolution are discussed in
\citet{ymaf05}, and effects that appear only in multi-D in Meakin,
Young, \& Arnett (submitted). These uncertainties in the initial
conditions exacerbate the unknowns of the explosion mechanism itself.

In spite of this, by taking advantage of the constraints we have on
the progenitor of Cas A, we can now use additional observations to
better understand the supernova explosion mechanism.  The explosion
energy of the Cas A supernova is predicted to be at the high end of
the range expected for normal supernovae: $2-4\times10^{51}$\,erg
(e.g. \cite{che03}).

\begin{figure}
\figurenum{4}
\plotone{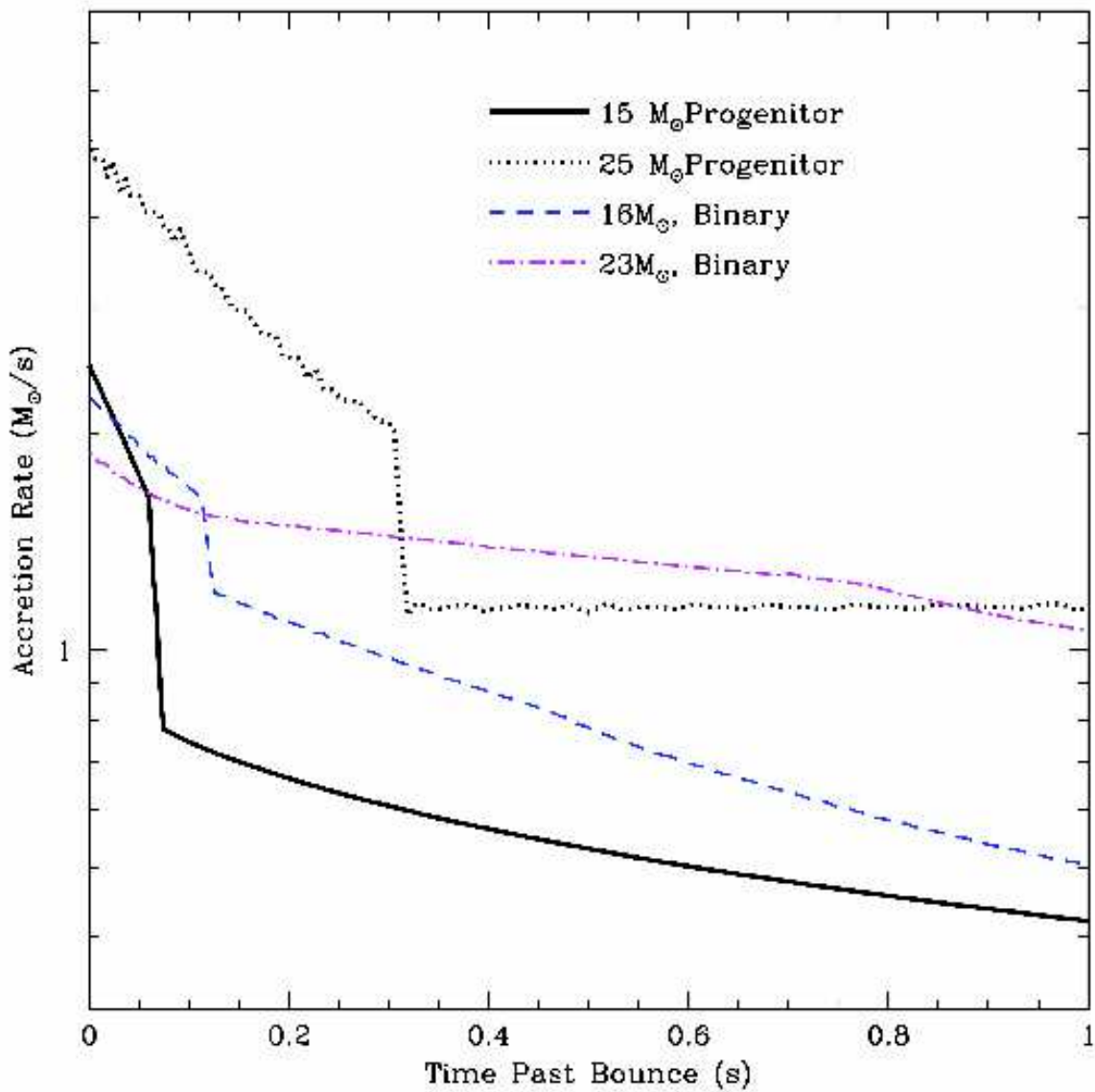}
\caption{Accretion rate vs. time since bounce for the 16 and 23 \sol
binary models from this paper and 15 and 25 \sol single stars from
\citet{fk01}. The time of the sharp drop determines the timing and the
accretion rate the energy of the explosion. Energetic explosions are
allowable in the region bounded by the 15 and 25 \sol models. A binary
model between 16 and 23 \sol can satisfy the prediction of an
energetic explosion \citep{che03}. \label{mdot} }
\end{figure}

Fryer \& Kalogera (2001) examine how supernova explosion energy varies
with mass. They argued that the supernova explosion energy peaked for
15M$_\odot$ progenitors.  At 23M$_\odot$, they predicted an explosion
energy of less than $10^{51}$\,erg. The progenitors we have used in
this paper are very different than the progenitors used in the work by
\citet{fk01}.  Those progenitors assumed single stars with {\it no}
mass loss, and the code used more approximate prescriptions for
nuclear burning and convection.  Instead of comparing star mass, it is
better to compare the fate of stars with specific structures.
Figure~\ref{mdot} compares the accretion rate onto the outer part of
the convective region during collapse (\citet{fryer99} describes why
this particular structure information is ideal in comparing
progenitors) for our 16 and 23M$_\odot$ stars against the 15 and
25M$_\odot$ from \citet{fk01}.  Already, knowing that a strong
explosion can be produced for stars with structures lying between our
16 and 23M$_\odot$ binary progenitors provides some of the most
accurate limits on the supernova explosion mechanism currently
available.  As we restrict our range of possible progenitors and place
stronger constraints on the explosion energy and yields, we provide a
very stringent test for any core-collapse code.  In this sense, Cas A
now can play an active role in validating core-collapse codes.
\clearpage
\clearpage
\acknowledgements This work was funded in part under the auspices of
the U.S.\ Dept.\ of Energy, and supported by its contract
W-7405-ENG-36 to Los Alamos National Laboratory, by a DOE SciDAC grant
DE-FC02-01ER41176, an NNSA ASC grant, and a subcontract to the ASCI
FLASH Center at the University of Chicago. BV thanks the Eugene
McDermott Scholarship Program for support during the summer of 2005 at
Los Alamos National Laboratory when this work was completed.

\clearpage
\begin{deluxetable}{lcccccccc}
\tablewidth{0pt}
\tablecaption{Explosion Simulations\label{tab:exp-sims}}
\tablehead{
  \colhead{Simulation}
& \colhead{M$_{\rm Prog}$}
& \colhead{Energy}
& \colhead{Bin.}
& \colhead{Asym.} 
& \colhead{M$_{\rm Rem}$} 
& \colhead{M$_{\rm Ejecta}$} 
& \colhead{$^{44}$Ti Yield} 
& \colhead{$^{56}$Ni Yield} \\

& \colhead{(\sol)}
& \colhead{$10^{51}$\,erg}
& \colhead{}
& \colhead{Jet2\tablenotemark{a}}
& \colhead{(\sol)}
& \colhead{(\sol)}
& \colhead{(\sol)}
& \colhead{(\sol)}

}
\startdata

M40E7.6 & 40 & 7.6 & N & N & 1.75 & 6.0 & $7.5\times10^{-5}$ & 0.33 \\
M23E0.8 & 23 & 0.8 & N & N & 5.4 & 7.5 & $<10^{-5}$ &  $<10^{-5}$ \\
M23E2.3 & 23 & 2.3 & N & N & 4.6 & 8.3 & $1.2\times10^{-5}$ & $2.6\times10^{-4}$ \\
M23E2.3A & 23 & 2.3 & N & Y & 5.5 & 7.4 & $1.8\times10^{-4}$ &  $0.19$ \\
M23E1.1Bin & 23 & 1.1 & Y & N & 2.6 & 3.6 & $1.2\times10^{-5}$ & $2.6\times10^{-4}$ \\
M23E1.1BinA & 23 & 1.1 & Y & Y & 3.2 & 3.0 & $1.6\times10^{-5}$ & $0.02$ \\
M23E2.0Bin & 23 & 2.0 & Y & N & 2.3 & 3.9 & $5.7\times10^{-5}$ & $0.055$ \\
M23E2.0BinA & 23 & 2.0 & Y & Y & 2.6 & 3.6 & $8.0\times10^{-5}$ & $0.075$ \\
M16E1.3Bin & 16 & 1.3 & Y & N & 1.8 & 3.25 & $<10^{-5}$ & $<10^{-5}$ \\
M16E1.1BinA & 16 & 1.12& Y & Y & 1.85 & 3.2 & $<10^{-5}$ & $<10^{-5}$\\
M16E3.1Bin & 16 & 3.1 & Y & N & 1.18 & 3.87 & $1.2\times 10^{-5}$ & 0.15 \\ 
M16E3.1BinA & 16 & 3.1 & Y & Y & 1.19 & 3.86 & $1.2\times 10^{-5}$& 0.15\\
\enddata
\tablenotetext{a}{See \citet{HFW04} for details.}
\end{deluxetable}
\clearpage

\begin{deluxetable}{lccccc}
\tablewidth{0pt}
\tablecaption{Simulation vs. Constraints\label{tab:obsvsim}}
\tablehead{
  \colhead{Simulation}
& \colhead{Nitrogen}
& \colhead{Ejecta}
& \colhead{Remnant}
& \colhead{$^{44}$Ti} 
& \colhead{$^{56}$Ni}\\

& \colhead{Clumps}
& \colhead{Mass}
& \colhead{Mass}
& \colhead{Yield}
& \colhead{Yield}

}
\startdata

M40E7.6 & N & N & Y & Y & Y? \\
M23E0.8 & N & N & N & N & N \\
M23E2.3 & N & N & N & N & N \\
M23E2.3A & N & N & N & Y & Y \\
M23E1.1Bin & Y & Y & Y? & N & N \\
M23E1.1BinA & Y & Y & N & N & Y \\
M23E2.0Bin & Y & Y & Y? & Y? & Y \\
M23E2.0BinA & Y & Y & Y? & Y & Y \\
M16E1.3Bin & Y & Y & Y & N & N \\
M16E1.1BinA & Y & Y & Y & N & N\\
M16E3.1Bin & Y & Y & Y & N & Y \\ 
M16E3.1BinA & Y & Y & Y & N & Y\\
\enddata
\end{deluxetable}
\clearpage


\end{document}